\documentclass[aps,prc,groupedaddress,showpacs,manuscript]{revtex4}

\usepackage{graphicx}

\begin{document}

\title{Pion freeze-out as seen through HBT correlations in heavy ion collisions from FAIR/AGS to RHIC energies}

\author {Qingfeng Li,$\, ^{1}$\footnote{E-mail address: liqf@fias.uni-frankfurt.de}
Marcus Bleicher,$\, ^{2}$ and Horst St\"{o}cker$\, ^{1,2}$}
\address{
1) Frankfurt Institute for Advanced Studies (FIAS), Johann Wolfgang Goethe-Universit\"{a}t, Max-von-Laue-Str.\ 1, D-60438 Frankfurt am Main, Germany\\
2) Institut f\"{u}r Theoretische Physik, Johann Wolfgang Goethe-Universit\"{a}t, Max-von-Laue-Str.\ 1, D-60438 Frankfurt am Main, Germany\\
 }


\begin{abstract}
We perform a systematic analysis of several HBT parameters in heavy
ion collisions from $E_{\rm beam}=2$~AGeV to $\sqrt {s_{\rm
NN}}=200$~GeV within the UrQMD transport approach and compare the
results to experimental data where available. We find that the
'lifetime' of the emission source as calculated from $\tau \sim
\sqrt{R_O^{2}-R_S^{2}}$, is larger than the experimentally observed
values at all investigated energies. The calculated volume of the
pion source ($V_f$) is found to increase monotonously with
increasing beam energy and the experimentally observed decrease of
the measured $V_f$ at AGS is not seen. Finally, we calculate the
mean free path $\lambda_f=0.5-1$~fm of pions at freeze-out and find
a good description of the experimental data above the AGS energy
region, supporting the suggestion of a universal kinetic decoupling
criterion up to the highest RHIC energies.
\end{abstract}


\pacs{25.75.Gz,25.75.Dw,24.10.Lx} \maketitle

In order to create the theoretically predicted {\it deconfined}
phase of Quantum Chromo-dynamics (QCD) heavy ions have been
collided with energies from less than $\sqrt s \sim 2.5$~GeV (SIS/FAIR energy regime), $2.5- 20$~GeV
(FAIR/AGS and SPS) up to $20 - 200$ GeV (RHIC). Indeed, it seems that some nontrivial signals
- such as charmonium suppression, relative strangeness enhancement, etc. -
of the (phase) transition to the deconfined phase have been observed in heavy ion collisions
(HICs) at SPS energies
\cite{Matsui:1986dk,Soff:1999et,Dumitru:2001xa,Gazdzicki:2006fy,Heinz:2006ur,Torrieri:2007qy}.
Additional information about the matter created in such collisions can be obtained from
the investigation of the space-time structure of the particle emission source (the region of homogeneity).
The established tool to extract this information is known as Femtoscopy \cite{Lisa:2005dd} or originally as
Hanbury-Brown-Twiss interferometry (HBT) \cite{HBT54,Goldhaber60,Bauer:1993wq}.
Experimentally this technique is quite often used
to extract the information on the spatio-temporal evolution
of the particle source, which has been scanned thoroughly by several
separate experimental collaborations over the whole discussed energy region
\cite{Lisa:2000hw,Ahle:2002mi,Kniege:2004pt,Kniege:2006in,
Adamova:2002wi,Back:2004ug,Adler:2001zd,Adcox:2002uc,Adler:2004rq,
Adams:2004yc,Antinori:2001yi,Antinori:2007jr}. However, so far the
measured excitation function of HBT parameters shows no {\it
obvious} discontinuities within the large span of explored beam energies
\cite{Lisa:2005dd}.

A comprehensive theoretical investigation on the excitation function
of the HBT parameters is thus highly required but still absent so
far \cite{Lisa:2005dd}. Recently, based on the Ultra-relativistic
Quantum Molecular Dynamics (UrQMD, v2.2) transport model (employing
hadronic and string degrees of freedom) (for details, the reader is
referred to Refs. \cite{Bass98,Bleicher99,Bra04,Zhu:2005qa}) and the
program CRAB (v3.0$\beta$) \cite{Koonin:1977fh,
Pratt:1994uf,Pratthome}, we have investigated the transverse
momentum, system-size, centrality, and rapidity dependence of the
HBT parameters $R_L$, $R_O$, $R_S$ (dubbed as HBT radii or Pratt
radii), and the cross term $R_{OL}$ of pion source at  AGS
\cite{lqf20063}, SPS \cite{lqf20062} and RHIC \cite{lqf2006}
energies, respectively. In general, the calculations are satisfying
and well in line with the experimental data although discrepancies
are not negligible. Such as, I), the calculated $R_L$ and $R_S$
values for Au+Au collisions at low AGS energies are visibly smaller
than the data if the default UrQMD version 2.2 (cascade mode) is
adopted. II), the HBT-'puzzle' with respect to the 'duration time'
of the pion source, is present at all energies. In order to
understand the origin of this HBT-'puzzle', some efforts were made,
however, a complete understanding of this effect is still lacking.

In this paper, we present pion interferometry results on  the "duration-time"
related quantity $\sqrt{R_O^{2}-R_S^{2}}$, the freeze-out volume $V_f$,
and derive the mean free path $\lambda_f$ of pions at freeze-out. The analysis is based
on the comprehensive comparison of the excitation function of calculated HBT radii $R_L$,
$R_O$, and $R_S$ at small transverse momenta with data, in the
FAIR/AGS, SPS and RHIC energy regime.
The standard UrQMD v2.2 in cascade mode is employed throughout
this paper to serve as a benchmark for further discussions\footnote{We
have found \cite{lqf20063} the treatment of the mass-dependent
resonance lifetimes can better describe the HBT radii for AGS
energies (at small transverse momenta), meanwhile, the HBT
time-related puzzle can be better understood with the consideration
of a potential interaction. These effects are, however, not taken
into account in this paper for the sake of clarity.}.

To calculate the two-particle correlator, the CRAB program is based
on the formula:
\begin{equation}
C({\bf k},{\bf q}) = \frac {\int d^4x_1 d^4x_2 g(x_1,{\bf p_1})
g(x_2,{\bf p_2}) |\phi({\bf q}, {\bf r})|^2} {{\int d^4x_1
g(x_1,{\bf p_1})} {\int d^4x_2 g(x_2,{\bf p_2})}}. \label{cpq}
\end{equation}
Here $g(x,{\bf p})$ is the probability for emitting a particle with
momentum ${\bf p}$ from the space-time point $x = ({\bf r}, t)$.
$\phi({\bf q}, {\bf r})$ is the relative two-particle wave function
with ${\bf r}$ being their relative position.
$\bf{q}=\bf{p}_2-\bf{p}_1$ and
$\textbf{k}=(\textbf{p}_{1}+\textbf{p}_{2})/2$ are the relative
momentum and the average momentum of the two particles. Due to the
underlying quantum statistics, this correlator is larger than unit
at small $q$ for bosons and can be fitted approximately by a
Gaussian form. Using Pratt's three-dimensional convention (the LCMS
system), the standard parametrisation of the correlation function in
Gaussian form reads
\begin{equation}
C(q_O,q_S,q_L)=1+\lambda
{\rm exp}(-R_L^2q_L^2-R_O^2q_O^2-R_S^2q_S^2-2R_{OL}^2q_Oq_L). \label{fit1}
\end{equation}
Here $q_i$ and $R_i$ are the components of the pair momentum
difference $\bf{q}$ and the homogeneity length (HBT radii) in the
$i$ direction, respectively. The pre-factor $\lambda$ is the incoherence parameter
and lies between 0 (complete coherence) and 1 (complete
incoherence) in realistic HICs. The term $R_{OL}^{2}$ is called
cross-term and vanishes at mid-rapidity for symmetric systems,
while it deviates from zero at large rapidities \cite{Chapman:1994yv,Wiedemann:1999qn,lqf20062}.

We compare our calculations of the Pratt parameters of the pion source with experimental data
for the following central collisions of heavy nuclei:
\begin{enumerate}
\item
Au+Au at the AGS beam energies $E_b=2$, $4$, $6$, and $8$A GeV
($<11\%$ of the total cross section $\sigma_T$), a rapidity cut
$|Y_{cm}|<0.5$ ($Y_{cm}=\frac{1}{2}{\rm
log}(\frac{E_{cm}+p_\parallel}{E_{cm}-p_\parallel})$, $E_{cm}$ and
$p_\parallel$ are the energy and longitudinal momentum of the pion
meson in the center-of-mass system) is employed. The
experimental (E895) data are taken from \cite{Lisa:2000hw}.

\item
Au+Au at the AGS beam energy $11.6$A GeV (the $<5\%$ most
central collisions), a rapidity cut $|Y_{cm}|<0.5$ is employed. The
experimental (E802) data are taken from \cite{Ahle:2002mi}.

\item
Pb+Pb at the SPS beam energies $E_b=20$, $30$, $40$, $80$, and
$160$A GeV ($<7.2\%\sigma_T$ of most central collisions), a
pion-pair rapidity cut $|Y_{\pi\pi}|<0.5$
($Y_{\pi\pi}=\frac{1}{2}{\rm log}(\frac{E_1+E_2+p_{\parallel
1}+p_{\parallel 2}}{E_1+E_2-p_{\parallel 1}-p_{\parallel 2}})$ is
the pair rapidity with pion energies $E_1$ and $E_2$ and
longitudinal momenta $p_{\parallel 1}$ and $p_{\parallel 2}$ in the
center of mass system) is employed. The experimental (NA49) data are
taken from \cite{Kniege:2004pt,Kniege:2006in}.

\item
Pb+Au at the SPS beam energies $E_b=40$, $80$, and $160$A GeV
(the $<5\%$ most central collisions), the pion-pair rapidity cut
$Y_{\pi\pi}=-0.25\sim 0.25$, $-0.5\sim 0$, and $-1.0\sim -0.5$ are
chosen. The experimental (CERES) data
are taken from \cite{Adamova:2002wi}.

\item
Au+Au at the RHIC nucleon-nucleon center-of-mass energies
$\sqrt{s_{NN}}=30$ ($<15\%\sigma_T$), $62.4$ ($<15\%\sigma_T$),
$130$ ($<10\%\sigma_T$), and $200$ GeV ($<5\%\sigma_T$). Here a
pseudo-rapidity cut $|\eta_{cm}|<0.5$ ($\eta_{cm}=\frac{1}{2}{\rm
log}(\frac{p+p_\parallel}{p-p_\parallel})$, ($p$ is the momentum of the
pion) is employed. The experimental (PHOBOS, STAR, and PHENIX)
data are taken from
\cite{Back:2004ug,Adler:2001zd,Adcox:2002uc,Adler:2004rq,Adams:2004yc}.
\end{enumerate}

\begin{figure}
\includegraphics[angle=0,width=0.8\textwidth]{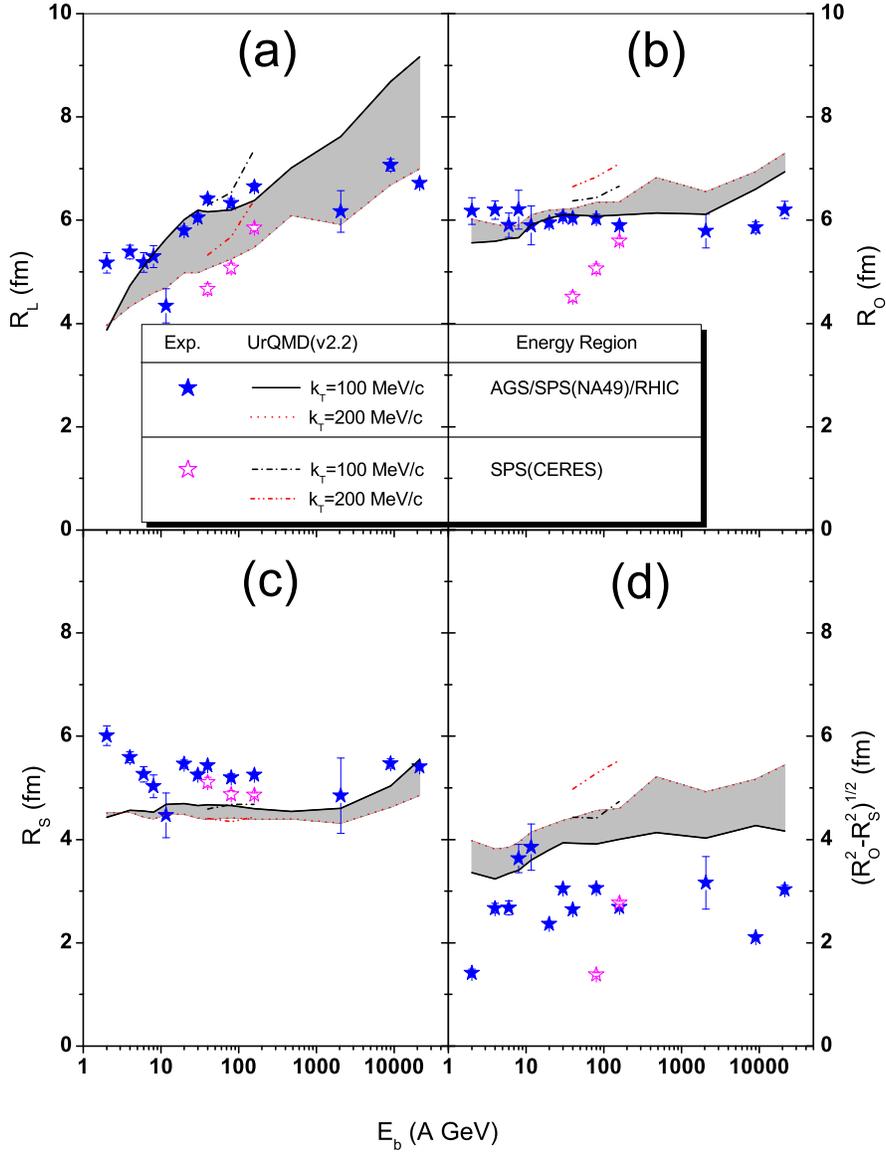}
\caption{(Color Online) Excitation function of HBT radii $R_L$ [in
(a)], $R_O$ [(b)], $R_S$ [(c)], and the quantity
$\sqrt{R_O^{2}-R_S^{2}}$ [(d)]. The calculations are shown at
$k_T=100\pm 50$~MeV (full line) and $200\pm 50$~MeV$/c$ (dotted
line), respectively.  The gray areas between the $k_T=100$~MeV and
$k_T=200$~MeV lines are shown for better visibility. The data are at
$k_T\sim 150$MeV$/c$ for reactions at $E_b=2,4,6,8 A$~GeV (AGS-E895)
\cite{Lisa:2000hw} and $20,30, 40, 80, 160 A$~GeV (SPS-NA49)
\cite{Kniege:2006in}, at $k_T\sim 170$MeV$/c$ for reaction at
$\sqrt{s_{NN}}=130$ GeV \cite{Adler:2001zd,Adcox:2002uc}, at
$k_T\sim 200$MeV$/c$ for reactions at $E_b=11.6$A GeV (AGS-E802)
\cite{Ahle:2002mi}, $E_b=40,80,160 A$~GeV (SPS-CERES)
\cite{Adamova:2002wi}, and $\sqrt{s_{NN}}=62.4,200$ GeV
\cite{Back:2004ug,Adler:2004rq,Adams:2004yc}.} \label{fig1}
\end{figure}

Fig.\ \ref{fig1} shows the excitation function of the calculated HBT
radii $R_L$ [in (a)], $R_O$ [(b)], $R_S$ [(c)], and the
duration-time related quantity $\sqrt{R_O^{2}-R_S^{2}}$ [(d)] at
$k_T=100$~MeV (full lines, black) and $200$~MeV (dotted lines, red).
The experimental data within this transverse momentum region are
shown for comparison. Since the experimental data from NA49
\cite{Kniege:2004pt,Kniege:2006in} and from CERES
\cite{Adamova:2002wi} collaborations overlap at beam energies $40$,
$80$, and $160$A GeV, we show the calculations and data with respect
to CERES energies separately as dashed-dotted lines and open
symbols.

The calculated $R_L$ (Fig.\ \ref{fig1}(a)) increases faster than
$R_O$ and $R_S$ with increasing energies, which is also observed in
data. Meanwhile, with increasing beam energies, the splitting of
$R_L$ with different $k_T$ becomes stronger, which can be attributed to the
flow-dominated freeze-out scenario. This observation is also in line with
previous results \cite{lqf2006}, where it was found that the decrease of $R_L$ with increasing
$k_T$ at RHIC energies is stronger for central reactions than for peripheral
ones. It should also be noted that the small $k_T$ behaviour of the correlation function
is also affected by some other factors, such as the decay of
resonances, potential interactions and/or the treatment of resonance life-times and widths
in the medium.

Fig.\ \ref{fig1} (b) shows the  calculations on
$R_O$ in comparison to the experimental data. Here we find that model calculation in
data agree fairly well at AGS and SPS-NA49 but deviations from data are observable at RHIC energies.
The comparison to the CERES data (with the appropriate cuts) however shows a rather
large discrepancy between calculation and data.
This deviations was also report in a previous study \cite{lqf20062} and might hint to
systematic differences (apart from different experimental centralities and cuts)
between the NA49 analysis and the results measured by the
CERES collaboration. The source of this difference remains therefore unclear. Recently, the NA57 collaboration published the $K_T$ dependence of the HBT radii (with $K_T$ up to $1.2$ GeV$/c$) in Pb+Pb collisions at $40$A GeV, and it was found that the NA57-data, especially the $R_O$, are in line with the NA49-data \cite{Antinori:2007jr}.

In Fig.\ \ref{fig1} (c), we notice that the calculated sideward
radii $R_S$ are in qualitative agreement with the data but seem to
be 15\% smaller for almost all energies. Furthermore, the increase
of the measured $R_S$ at low AGS energies can not be reproduced
\cite{lqf20063} and might be due to the omission of potential
interactions that gain importance at low beam energies. Detailed
investigations will be presented in a future publication. The
excitation function of the HBT radii from several systems (from
light to heavy) inspired by the NA49-future collaboration
\cite{Gazdzicki:2006fy} might also provide new insights into this
problem, and predictions are in progress.

The HBT duration time ''puzzle'', i.e. the fact of the theoretical quantity
$\sqrt{R_O^{2}-R_S^{2}}$ being larger than extracted from the data, is present at all
investigated energies (see Fig.\ \ref{fig1} (d)): The calculated values
of $\sqrt{R_O^{2}-R_S^{2}}$ are about $3.5\sim 5$~fm while the
measured ones are $1.5\sim 4$~fm. Many efforts have been put forward over the last years
to clarify this issue
\cite{Lin:2002gc,Cramer:2004ih,Humanic:2005ye,Pratt:2005hn,Pratt:2005bt,lqf2006,lqf20062,lqf20063}.
E.g., in our previous works \cite{lqf20062,lqf20063}, we suggested that
a sizeable amount of interactions between particles at the early stage of the reaction
either on the mean field  and/or (non-perturbative) partonic level seem to be relevant to understand
the phenomenon. In fact, all straightforward cascade transport
approaches (also those with partonic interactions with perturbative QCD cross sections)
fail to describe the quantity $\sqrt{R_O^{2}-R_S^{2}}$, as well
as the elliptic flow \cite{Petersen:2006vm}, over the whole energy
range.
\begin{figure}
\includegraphics[angle=0,width=0.6\textwidth]{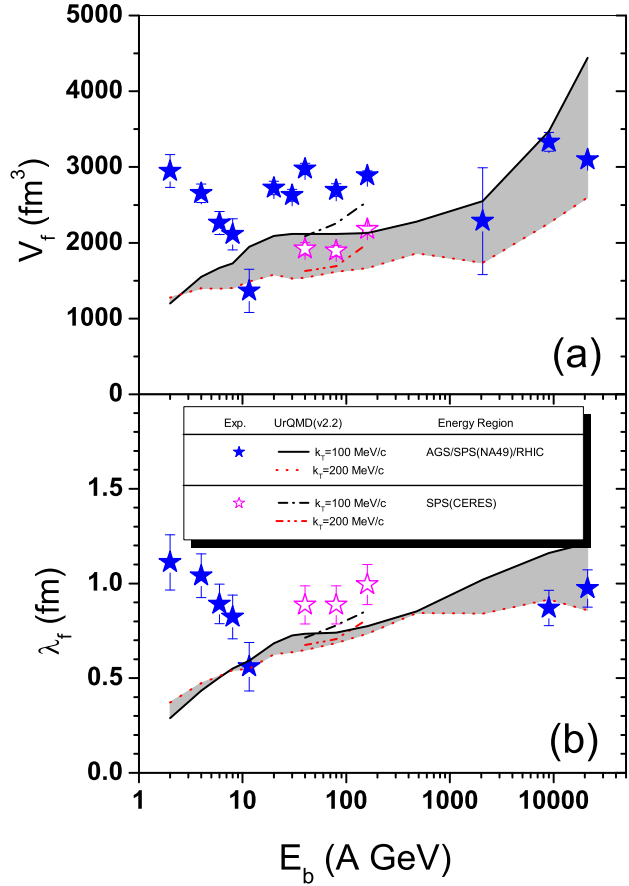}
\caption{(Color Online) (a): Excitation function of the pion freeze-out
volume $V_f$ (according to Eq. \ref{volume1}) at transverse
momenta between $k_T=100$~MeV and $200$~MeV (gray area), compared with data in this
$k_T$-region. (b): Excitation function of the mean free path
$\lambda_f$ of pions at freeze-out (according to Eq. \ref{lam1}) at
the same transverse momenta. The experimental value for
$\lambda_f$ at $\sqrt{s_{NN}}=200$~GeV is obtained with the help of
recent $dN/dy$ data in \cite{Adler:2003cb}, at all other energies the $\lambda_f$
data are taken from \cite{Adamova:2002ff}.} \label{fig2}
\end{figure}

Fig.\ \ref{fig2} (a) shows the excitation function of the pion
source volume $V_f$ at freeze-out, calculated as \cite{Adamova:2002ff}
\begin{equation}
V_f=(2\pi)^\frac{3}{2}R_LR_S^2. \label{volume1}
\end{equation}
Note that the radius $R_O$ is not considered to calculate the pion
freeze-out volume since it contains the contribution of the temporal
extent of the pion source. Fig.\ \ref{fig2} (a) shows clearly that
the UrQMD cascade calculations do provide a reasonable freeze-out
volume for the pion source at RHIC energies. At SPS energies, the
agreement is fine with CERES data while it slightly underpredicts
those of NA49. Towards even lower energies, the model underpredicts
the measured freeze-out volume due to the omission of the strong
interaction potential and other in-medium effects. E.g. at $E_b=2A$
GeV, the measured $V_f$ is  about $2-3$ times larger than calculated
value. As studied in \cite{lqf20063}, a mass-dependent lifetime of
resonances accounts for an improvement of the HBT-radii at small
$k_T$ and hence reproduce the data better. From the discussions
above, it is clear that the major part of this difference is related
to the sideward radius $R_S$.

We are now ready to estimate the mean free path $\lambda_f$ of the pions at freeze-out
from the following expression \cite{Adamova:2002ff}
\begin{equation}
\lambda_f=\frac{V_f}{N\sigma}=\frac{V_f}{N_N\sigma_{\pi
N}+N_{\pi}\sigma_{\pi\pi}} \label{lam1}.
\end{equation}
with the averaged pion-nucleon cross section $\sigma_{N\pi}=72$~mb
and the averaged pion-pion cross section $\sigma_{\pi\pi}=13$~mb
(Note that within the present model calculations these values are
slightly energy dependent. However, here we have adopted the
explicit numbers from Ref. \cite{Adamova:2002ff} to compare to the
results presented there). The nucleon and pion multiplicities $N_N$
and $N_{\pi}$ are calculated as

\begin{equation}
N_N=y_{th} \cdot \sqrt{2\pi} \cdot
\left.\frac{dN_{nucleons}}{dy}\right.|_{y_{mid}}, \label{Nn1}
\end{equation}
and
\begin{equation}
N_\pi=y_{th} \cdot \sqrt{2\pi} \cdot
\left.\frac{dN_{pions}}{dy}\right|_{y_{mid}} \label{Npi1}.
\end{equation}
using the assumption of a thermal equilibrated system at freeze-out with
a temperature $T_f=120$ MeV. Here, $y_{th}$ is the estimated
thermal homogeneity scale in rapidity at a certain $k_T$ and $T_f$, and is
given by the  expressions: $y_{th}={\rm
arctanh}(\langle\beta_{th}\rangle)$, with
$\langle\beta_{th}\rangle=\sqrt{1+\langle\gamma\rangle^{2}}/\langle\gamma\rangle$ and
$\langle\gamma\rangle=1+1/3\left(K_1(m_T/T_f)/K_2(m_T/T_f)-1\right) +T_f/m_T$. Here $K_n(z)$ is the modified
Bessel function of order $n$ and $m_T=\sqrt{m_\pi^2+k_T^2}$. At $k_T=100$~MeV and
$200$~MeV, the calculated homogeneity lengths in rapidity $y_{th}$ are
0.98 and 0.81, respectively. $dN/dy|_{y_{mid}}$ is the rapidity density of
pion (nucleons) at mid-rapidity\footnote{In Ref. \cite{Adamova:2002ff}, the
experimental $N_{nucleons}$ is set to two times of total amount of
emitted proton and anti-proton yields and $N_{pions}$ is three times
of negatively charged pion number. In our calculations, $N_{nucleons}$ represents
the sum of protons and neutrons, as well as  anti-nucleons.
Note that this introduces a small difference into the analysis
due to iso-spin effects, especially at very low energies.}.
Recent calculations using the present UrQMD transport
model \cite{Bra04}, have shown that the calculated
pion and nucleon yields are reasonably in
agreement with data.

Fig.\ \ref{fig2} (b) shows the excitation function of $\lambda_f$ of
pions at freeze-out. It is seen that the theoretical $\lambda_f$
value increases gradually from $\sim 0.5$ to $\sim 1$~fm from AGS to highest RHIC energies
with a weak dependence on $k_T$. The
experimental values of $\lambda_f$ are also between $0.5-1$~fm.
The calculated $\lambda_f$ below the AGS energy deviates from the data
due to the smaller calculated sideward radii (resulting in a too small freeze-out volume).

Certainly, the assumption of a constant kinetic freeze-out
temperature $T_f=120$~MeV in the calculations of $y_{th}$ over the whole energy
range is not justified, especially for the collisions at low beam energies.
By inserting an energy dependent temperature (assuming a kinetic freeze-out temperature of 70\% of the
chemical freeze-out temperature shown in \cite{Cleymans:2006qe})
into Eqs.\ \ref{Nn1} and \ref{Npi1}, one finds that at lower AGS energies,
the number of nucleons and pions in the pion source volume is
visibly reduced due to the decreased $T_f$. As a result
$\lambda_f$ increases and a flatter distribution of
$\lambda_f$ as a function of the beam energy is obtained in the present calculation.

The observation (both experimentally and theoretically) of a nearly
energy independent mean free path on the order of $0.7$~fm at pion
freeze-out is rather surprising. Physically it has been interpreted
as a rather large opaqueness of the pion source at break-up
\cite{Heiselberg:1997vh,Tomasik:1998qt}.

To summarize, we survey the excitation function of the HBT radii
$R_L$, $R_O$, and $R_S$, the quantity $\sqrt{R_O^{2}-R_S^{2}}$, the
volume $V_f$, and the mean free path $\lambda_f$ of pions at
freeze-out for heavy systems with energies ranging from lowest AGS
to the highest RHIC energies. Generally, the model calculations with
UrQMD v2.2 (cascade mode) are  in line with the data over the whole
inspected energy range. Although discrepancies especially in the
lower AGS energy region are found and have to be resolved.
Especially the HBT parameter $R_S$ and the HBT duration-time related
"puzzle" deserve further attention. Finally we re-interpret the
measured and calculated radii in terms of a mean free path for pions
and kinetic freeze-out. Here we find a nearly constant mean free
path for pions on the order of $\lambda_f=0.7$~fm indicating a
significant opaqueness of the source.

\section*{Acknowledgements}
We would like to thank S. Pratt for providing the CRAB program and
acknowledge support by the Frankfurt Center for Scientific Computing
(CSC). Q. Li thanks the Frankfurt Institute for Advanced Studies
(FIAS) for financial support. This work is partly supported by GSI,
BMBF, DFG and Volkswagenstiftung.

\end{document}